%% file: main.tex
\documentclass[lettersize,journal]{IEEEtran}

\usepackage{cite}
\usepackage{graphicx}
\usepackage{amsmath,amssymb,amsthm}
\usepackage{multirow}
\usepackage{acro} \acsetup{format/foreign = {\emph}}
\input{acronyms.tex}

\usepackage{cleveref}
\usepackage{subfigure}

\usepackage{soul,color}
\soulregister\ac7
\usepackage[none]{hyphenat}

\begin{document}

\title{Semantic-Functional Communications in Cyber-Physical Systems}


\author{
Pedro E. Gória Silva,
Pedro H. J. Nardelli,  \textit{Senior Member}, \textit{IEEE},
Arthur S. de Sena,  \textit{Member}, \textit{IEEE},
Harun Siljak, \textit{Senior Member}, \textit{IEEE},
Niko Nevaranta, 
Nicola Marchetti, \textit{Senior Member}, \textit{IEEE},
Rausley A. A. de Souza, \textit{Senior Member}, \textit{IEEE}
\thanks{Pedro Gória Silva, Pedro H. J. Nardelli, and Niko Nevaranta are with Lappeenranta--Lahti University of Technology, Finland, (email: pedro.goria.silva@lut.fi, pedro.nardelli@lut.fi, niko.nevaranta@lut.fi). 
P. H. J. Nardelli is also with the University of Oulu, Finland. 
Arthur S. de Sena is with the University of Oulu, Finland, and with the Technology Innovation Institute, United Arab Emirates (email: arthurssena@ieee.org).
Harun Siljak and Nicola Marchetti are with Trinity College Dublin, Ireland (email: harun.siljak@tcd.ie, nicola.marchetti@tcd.ie).
Rausley A. A. de Souza is with the National Institute of Telecommunications (Inatel), Santa Rita do Sapucaí 37540-000, Brazil. E-mail: rausley@inatel.br.
%
}
}


\maketitle

\begin{abstract}
This paper explores the use of semantic knowledge inherent in the \ac{CPS} under study in order to minimize the use of explicit communication, which refers to the use of physical radio resources to transmit potentially informative data.
It is assumed that the acquired data have a function in the system, usually related to its state estimation, which may trigger control actions.
We propose that a semantic-functional approach can leverage the semantic-enabled implicit communication while guaranteeing that the system maintains functionality under the required performance.
We illustrate the potential of this proposal through simulations of a swarm of drones jointly performing remote sensing in a given area.
Our numerical results demonstrate that the proposed method offers the best design option  regarding the ability to accomplish a previously established task---remote sensing in the addressed case---while minimizing the use of radio resources by controlling the trade-offs that jointly determine the \ac{CPS} performance and its effectiveness in the use of resources.
In this sense, we establish a fundamental relationship between energy, communication, and functionality considering a given end application.
\end{abstract}

\section{Introduction}
\IEEEPARstart{T}{he} fundamental limit of \textit{explicit} communication, i.e., when a physical medium is used to transfer data from one point to another, was determined by Shannon in his seminal work \textit{A Mathematical Theory of Communication} \cite{shannon1948mathematical}.
Channel capacity was then defined as the maximum rate at which data can be transferred with an error probability approaching zero, also proving the existence of a code that can achieve it.
Following the terminology of linguistics \cite{denham2012linguistics}, this refers to the \textit{syntax} of the communication problem (i.e., the explicit structure of ``words" that constitutes what is a ``well-formed sentence").
\textit{Semantics} is different: it refers to meaning (i.e., the signification of ``words").

By shifting the domain from syntax to semantics, the communication problem changes from optimizing the use of the physical medium to when and how to use it in the most effective way, considering both the meaning of the data to be transmitted and the use of the data at the end point.
In this case, the actual meaning of the data to be transmitted is an attribute of a given system, process, or phenomenon.
Following the conceptualization of  \cite{nardelli2022cyber}, we can say that the semantics is immanent from the \ac{CPS} in consideration, where data are acquired from a physical phenomenon to be then processed before being used in a decision-making process that determines the actuation; all types of data processing refer to the cyber domain.
Moreover, all data are acquired with certain goals in mind, like controlling an industrial process or monitoring a given environment.
This conceptualization is also supported in a broader perspective by \cite{kolchinsky2018semantic}, where it is stated that: \textit{Shannon information theory provides various measures of so-called syntactic
information, which reflect the amount of statistical correlation between systems.
By contrast, the concept of ‘semantic information’ refers to those
correlations which carry significance or ‘meaning’ for a given system.}

A novel approach called \textit{semantic-functional communications} (\acs{SFC}) was recently proposed in \cite{silva2022semantic,silva2023enabling}.
The difference between the proposed \ac{SFC} and other communication system designs, is to design the physical layer in order to consider the ``idle states'' of given radio resources as informative to the receiver in semantic terms, and provide an efficient method of accessing the medium based on the semantic understanding of the data. 
In other words, by not explicitly transmitting any data, the receiver ``recovers'' the information about the predetermined state of the system under consideration. 
This was tested in a multiuser scenario, where different sensors need to transmit only alarm messages to a decision-maker node (receiver) that needs to (i) identify if a message was transmitted and (ii) by which sensor.
Such an initial case-specific solution motivated us to generalize this idea in order to characterize the fundamental limits of \ac{SFC}. 

This article focuses on the fundamental communication limits of transmission of informative semantic data to perform well-determined functions within \acp{CPS}.
The proposed approach consists of minimizing the use of explicit data transfer, i.e., maximizing ``implicit communications,'' by leveraging the ``informative silences'' of idle states.

\section{Background: Semantic communications}
\label{sec:Background}
Following the current trend, the technical literature, as a matter of fact, does not provide a precise or standardized delimitation for the concept of semantic or goal-oriented communications; it has dealt with this issue by distinguishing it from the syntactic (Shannon-like) approach to the communication problem.
In simple terms, \textit{syntactic} communications aim to establish a flow of information without taking into account any intrinsic quality of the data, while \textit{semantic} communications explore the (preliminary) meaningful knowledge of the data that entities may previously have, or are capable of acquiring.

\begin{figure*}[!ht]
 \centering
\includegraphics[width=1\textwidth]{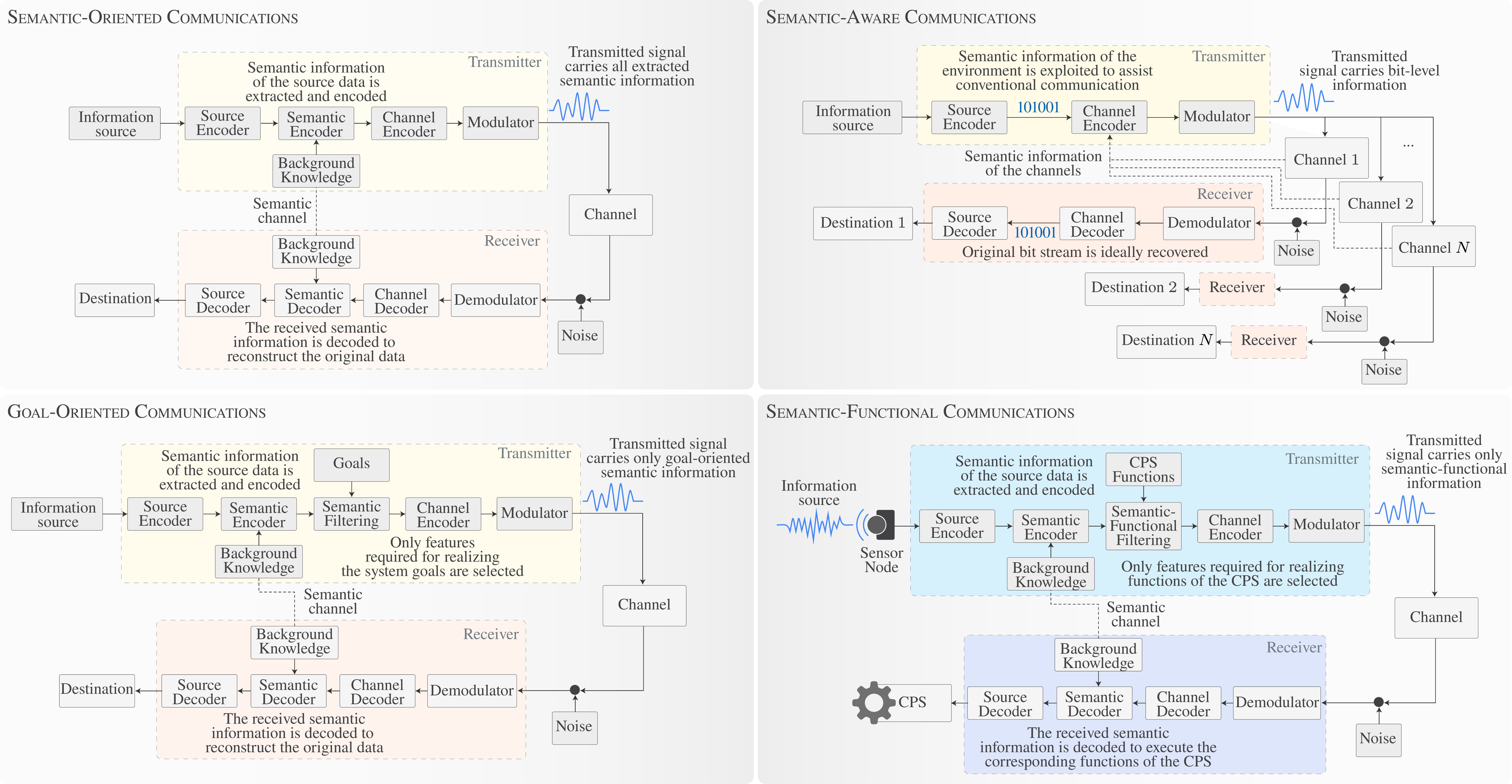}
 \caption{Comparison between different classes of semantic communications.}
  \label{fig_var}%
\end{figure*}

However, a sharper distinction between these two theoretical classes requires more attention. 
We will discuss how the current literature on wireless communication deals with the ``semantics'' based on three main categories: \textit{semantic-oriented communications}, \textit{goal-oriented communications}, and \textit{semantic-aware communications}. 
The reader is encouraged to peruse \cite{YangSemComSurvey} to thoroughly understand the most recent developments in semantic communication theory and the reasons behind such a classification.
Figure \ref{fig_var} presents a comparison between the different classes of semantic communication.
In the following, we will provide a brief overview of the topic.

\subsection{Semantic-oriented communications}
A semantic-oriented communication system bases its coding process on the quantification of semantic information, i.e., by primarily using the \textit{Theory of Strongly Semantic Information}~\cite{Floridi2004}, it aims to achieve semantic-level communications. 
One strand of this theory models the communication system through a fundamental type source that can make factual statements in propositional logic. 
The major distinction from Shannon's information theory is that logical probabilities guide the semantic information measure. 
These logical probabilities are defined by background knowledge instead of statistical probabilities. 
Furthermore, one can consider the prior knowledge of the destination about the source in the coding process. 
Therefore, the main difference of a semantic-oriented communication system from a traditional one is the data processing based on semantic characteristics. 

%
%
%
\subsection{Goal-oriented communications}
In the last decades, research efforts have been made to study and clarify the concept of \textit{goal-oriented communications} by defining \textit{goals} (or \textit{tasks}) in communication.
%
On the one hand, we have goals that capture the intent of communicating parties, called \textit{meta-goals}. 
On the other hand, we have goals that capture effects observed by a third party, named \textit{syntactic goals}. 
Based on these definitions, some papers indicate that communications systems can accomplish meta-goals even if the two parties do not share the same language (i.e., protocol). 
To make this feasible, the third party must be able to assess whether the meta-goals have been achieved.
%

In recent works on goal-oriented communications, the aim is to take advantage of the growth of computational power in conjunction with state-of-the-art \ac{AI} techniques. 
While the communication principle---enabling the parties to perform collaborative communication aiming at a common goal or task---remains unchanged, the optimization of task (the process performed by the parties in order to complete the task) is carried out by recently developed \ac{AI} algorithms.
One example is the task-oriented multiuser semantic communication system for multimodal data transmission proposed in \cite{Xie9653664}. 
There, the authors discuss how semantic transmission can be used to inquire information about images by correlating multimodal data from multiple users.
%


\subsection{Semantic-aware communications} 

The two classes we just discussed, semantic-oriented and goal-oriented communications, address a connection between two specific agents. 
Even though the network is made up of multi-agents, the peer-to-peer interaction works as a basis for completing a task (meta-goal).
In other words, one can plainly distinguish an explicit pair of agents (source and destination) according to communication purposes (meta-goal) for semantic-oriented and goal-oriented communications. 
In contrast, semantic-aware communication ties a task that has to be performed jointly by several agents to the communication system. 
In this way, multiple agents cooperate to complete a task in a centralized or distributed way through semantic-based communications. 
Semantic-aware communication differs from semantic-oriented and goal-oriented communications by approaching the semantic information as a result of agent behavior in the environment instead of extracting it from a data source. 

We can give the following example. 
A base station communicates with several users spread over a given area. 
The base station tries to predict the conditions, i.e., the semantic information, of the mobile radio channel between itself and all users. 
The base station sets the schedule of broadcasts based on its channel forecast.
%
Furthermore, there are case studies in the literature related to scenarios with multiple intelligent agents in which the semantic information extracted is specifically used to facilitate multi-agent cooperation, regardless of the task or goal of the agents themselves~\cite{YangSemComSurvey}.
%

\section{Semantic-functional communications for \acp{CPS}}\label{sec:SFC}
As briefly introduced in \cref{sec:Background}, the term \textit{semantics} has a broad range of meanings in the wireless communication literature, dealing with aspects like information extraction, encoding, and contextualization. 
Our use of the term refers to the understanding of raw data as information within a given context, system, or end application.
To avoid confusion, we define semantics as the outcome of a symbolic operation that assigns a purified (no misunderstanding) meaning in the cyber domain to data that have potential to be directly or indirectly acquired from the physical world. 
For a given \ac{CPS}, it is possible to further specify the semantics by determining the subset of all acquired data that support the \ac{CPS} to function within its operational limits; this subclass is named here as \textit{semantic-functional}. 
Therefore, by semantic-functional, we mean semantics that enables the \ac{CPS} to perform its function.

Although our definition of semantic-functional and the pragmatic information---at least as the recent literature propose it as in \cite{mason2023multi}---are somewhat similar, we decide to keep the term semantic-functional because of the origin of the term pragmatics.
The fields of linguistics and philosophy pioneered the definition of pragmatics, which can be briefly summarized as follows.
Pragmatics has to do with the relations of signs to those who use the signs~\cite{Morris1938}.
Therefore, pragmatics does not directly deal with the effect of the message on system functionality; in our understanding, the use of pragmatic information in the recent communication theory would be better named practical information.


Let us illustrate the previously introduced concept of semantic and semantic-functional. 
Assume that the temperature must be kept below a threshold by controlling the window in a closed room. 
The physical world, in this example, refers to the gases inside the room, window, walls, etc., while the temperature is an abstract concept that measures a statistical average of the level of agitation between molecules. 
By defining a scale (Kelvin, Celsius, etc.) for the mean agitation of molecules, we are, within our context, attributing semantics to a physical phenomenon and the data acquired therefrom. 
Furthermore, this purified concept of mean agitation is semantic-functional in our case because it enables the \ac{CPS} to operate. 
Note that one can obtain different degrees of freedom from our semantic-functional measurements. 
For example, one can measure the temperature with a few tenths of precision or acquire a binary variable that expresses temperature being above or below the threshold; the most suitable design option depends on how strict are the functional requirements of the \ac{CPS} in terms of desired room temperature, including the frequency of possible violations of such constraint. 
On the other hand, although a scale to measure the standard deviation of molecular agitation is semantic, it would not be semantic-functional because it does not have any role in the \ac{CPS} operation. 

Without loss of generality, let $a_i[k] \in \mathcal{A}$ be the $i$th attribute of a physical system $\Phi$ at a given (discrete) time $k \in \mathbb{N}^+$ and defined by a set of possible meaningful attributes $\mathcal{A}$, such as temperature and position.
%
The series of the attribute $a_i[k-t],\dots, a_{i}[k]$ $\forall i=\{1,\dots,N\}$ is the basis to determine the semantic-functional state $s_{k} \in \mathcal{S}$, where $t$ is the time dependence, and $\mathcal{S}$ contains the possible semantic-functional states that $\Phi$ can experience, e.g. whether an attribute is outside of its operating range; this is defined as a history-dependent mapping (a semantic operation) from $f: \mathcal{A}^{N\times (t-1)} \rightarrow \mathcal{S}$.
Note that both $a_i$ and $s_k$ are symbolically defined measurements, with the former being the data acquired from $\Phi$ and the latter being the semantic-functional representation of data in the cyber domain.

More specifically, let $X \in \mathcal{X}$ and $Y \in \mathcal{Y}$ represent the input and output messages of a communication channel, with $\mathcal{X}$ and $\mathcal{Y}$ corresponding to the input and output message sets, respectively.
Also, we let the estimation of $s_k$ in the decision-maker be denoted by $\hat{s}_k = r(Y)$, with $r: \mathbb{C} \rightarrow \mathcal{S}$ being a function to recover the semantic-functional state of $\Phi$.
Last, we have $X=m(s_k,s_{k-1},\dots,s_{k-d})$ with $m : \mathcal{S}^{d+1} \rightarrow \mathcal{X}$ being the transmission function that incorporates aspects of signal modulation, encoding, and medium access control, where $d$ is the time dependence.
It is noteworthy that non-transmission (silence or idle channel) is a valid state for $X$.

Let us assume an ideal case where $\Phi$ operates as expected (no noise or unexpected behavior).
For example, if $\Phi$ is a dynamical system that can be fully characterized over time by its equations of motion and initial states, it is possible (and reasonable) that this knowledge is shared between the transmitter and receiver sides \textit{before} the operation of the system.
In this case, it is unnecessary to monitor $\Phi$.
Now let us consider the case where the physical system is not ideal, and, indeed, diverges from the expected dynamics.
In this case, we would have $\textup{Pr}[\hat{s}_k \neq s_k] > 0$ without explicit communication. 
One can reduce $\textup{Pr}[\hat{s}_k \neq s_k]$ as much as desired by performing explicit communication.
However, we aim to minimize explicit communication---even though we have $\textup{Pr}[\hat{s}_k \neq s_k] > 0$---while keeping the physical system operating.

In this context, the challenges of semantic-functional communication are: (i) determining the semantic-functional states $s \in \mathcal{S}$ for the operation of the cyber-physical system; (ii) proposing an event-driven communication system in order to minimize the need for explicit communication; and (iii) defining what the acceptable level of errors that the \ac{SFC} requires is, so that the cyber-physical system is still functional considering the proposed data transmission scheme.

The mathematical formulation of one of the simplest possible \ac{SFC} problems can be stated as follows: the aim is to minimize the average occurrence of transmissions $\mathbb{E}[\Theta[k]]$ so that the performance indicator $\mathcal{V} < \delta$ is always met.
$\Theta : \mathcal{X} \rightarrow \{0,1\}$ is an indicator function that takes on the value 0 if the current semantic-functional state $s$ must not be transmitted, and 1 otherwise, $\mathcal{V}$ refers to a given performance indicator of the \ac{CPS} in consideration, and $\delta$ is the maximum acceptable value of $\mathcal{V}$ so that one can operate a given function of the \ac{CPS}.
\section{Illustrative example}\label{sec:Illus Example}
In this section, we illustrate the potential of the proposed solution considering a remote sensing task to be performed by a swarm of drones that need to be coordinated via wireless communications.
It is worth indicating that this scenario, which is motivated by \cite{qin2022coordination}, is used here as an example to demonstrate the benefits of the proposed SFC approach to control the trade-offs that ultimately determine the performance of this \ac{CPS}  operation.

\subsection{Scenario description}
\begin{figure*}
 \centering
 \includegraphics[width=1\textwidth]{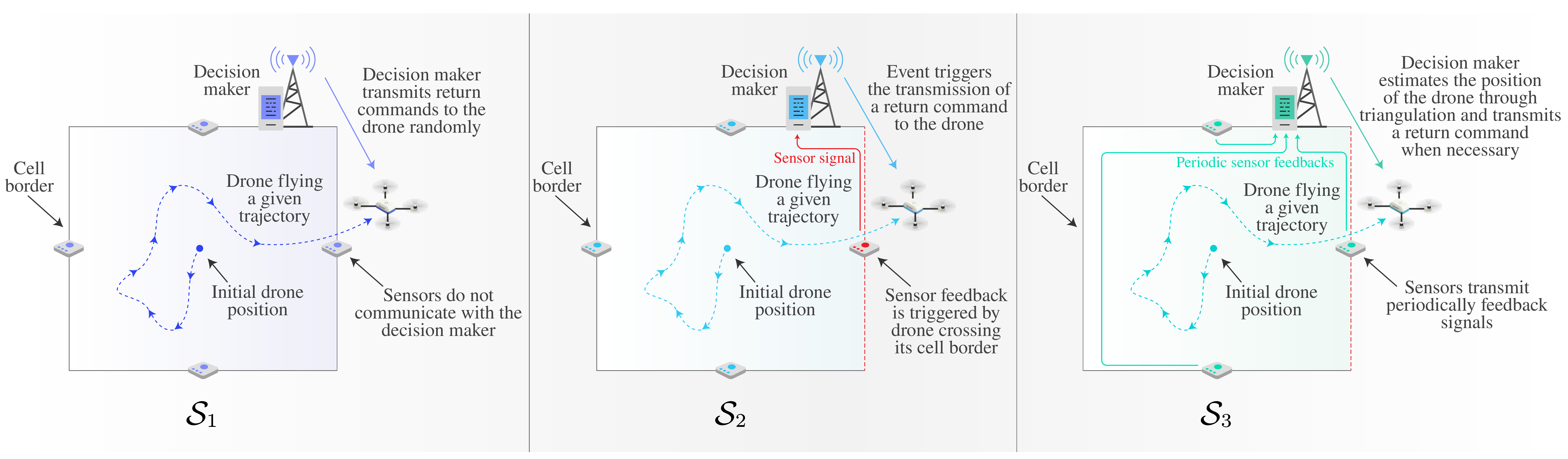}
 \caption{Simplified illustration of the three proposed solutions: $\mathcal{S}_{1}, \mathcal{S}_{2}$, and $\mathcal{S}_{3}$.}
  \label{fig:Positions}%
\end{figure*}

The target is to monitor a specific area by performing remote sensing by a swarm of nine autonomous drones, which are assumed to be able to satisfactorily monitor such an area. 
To this end, the area is divided into nine square cells, assigning one drone to each cell. 
Figure \ref{fig:Positions} outlines a simplified illustration of one cell for each of the three proposed approaches: $\mathcal{S}_{1}, \mathcal{S}_{2}$, and $\mathcal{S}_{3}$. 
The drones are programmed to perform a random walk and return to the central position of their coverage area when they receive a return command (or a track request) indicating that they are outside of their specific region. 
After returning to the center, the drone restarts its random walk.
We abstracted from our analysis the implications of the drone flight altitude, as this would result in unnecessary complexity for our explanation.

The purpose of our communication system is to enable the nine drones to stay within their respective areas using as few communication resources as possible (i.e., minimizing explicit communications), while controlling the performance indicator $\mathcal{V}_{\text{time}}$ and $\mathcal{V}_{\text{violation}}$ so that their value does not exceed its maximum acceptable upper bound. 
In our scenario, we have $\mathcal{V}_{\text{time}}=\{ t_1,\dots, t_9 \}$ and $\mathcal{V}_{\text{violation}}=\{ v_1,\dots, v_9 \}$, where $t_i$ and $v_i$ are the average time required for the $i$th drone to cover its respective area and the percentage of time the $i$th drone remained \textit{outside} its respective area (perimeter violation), respectively.
Note that the system allows, within certain limits, a drone to leave its area or overly concentrate navigation in the center of its coverage area. 
Both cases refer to inadequate coverage and lead to an increase in the average time needed to cover the entire area.
The remote sensing task (e.g., what is monitored, how and to whom the drones will report their measurements) is beyond the scope of this work.

We assume that the decision-maker can transmit the return command to drones through broadcast, error-free communication.
A series of protocols and modulation standards can be listed as candidates for such communication; however, it is enough for our explanation to assume that the drones can receive the return command with negligible delay, and that the decision-maker has unlimited access to the control channel of the drones.
In addition, we have separate channels for control and sensor communication.
\subsection{Semantic-functional communication}
\label{ssec:SFC cases}
Our first challenge is to determine the semantic-functional states for the operation of the cyber-physical system under $\mathcal{V}_{\text{time}}<\delta_{\text{time}}$ and $\mathcal{V}_{\text{violation}}<\delta_{\text{violation}}$.
Note that $\delta_{\text{time}}$ and $\delta_{\text{violation}}$ can represent either a physical constraint or a quantitative assessment of system functionality. 
For example, $\delta_{\text{time}}$ can be the maximum battery time of a drone.
We propose three solutions as follows.
\begin{enumerate}
    \item $\mathcal{S}_1 = \emptyset$, no system state is considered. This is an extreme case where the system is not monitored and there is no communication from sensors to the decision-maker. However, the decision-maker continues to act on the drones by randomly sending a return command to each drone.
    \item $\mathcal{S}_2$ is constituted by the state that the system assumes regarding whether the drones are within their respective coverage areas. For example, a state would be all drones inside their respective subareas, another semantic-functional state would be to have just the first drone outside its respective subarea, and so on.
    \item $\mathcal{S}_3$ has all possible combinations of the position (coordinates) of each drone. In other words, $\mathcal{S}_3$ refers to the drone positions for a given accuracy scale. For instance, a state would be all drones in their respective starting positions (center of subarea).
\end{enumerate}

Let us propose a communication system in order to minimize the need for explicit communication from sensors to the decision-maker.
Among some practical solutions to support system operation, we can highlight two: \textit{event-triggered communication} for $\mathcal{S}_2$ and \textit{periodic communication} for $\mathcal{S}_3$.

With regard to $\mathcal{S}_2$, an excellent approach would be to adopt event-triggered transmissions; in this way, the sensor nodes only transmit when the drone crosses the border.
For each sensor node, two events are possible: a drone crosses the border from right to left (or from top to bottom) or vice versa.
As shown in \cite{silva2022semantic} for event-triggered transmissions, the semantic-functional communication technique overcomes orthogonal and random protocols for media access control regarding $\textup{Pr}[\hat{s}_k \neq s_k]$. 
Furthermore, this transmission scheme incorporates the process of modulation and encoding.
Note that the semantic-functional communication technique of \cite{silva2022semantic} addresses how the sensors access the medium, but not the temporal dependence required by transmission function $m$.

The technique proposed in \cite{silva2022semantic} ensures that the decision-maker is aware of the event even when collisions (two or more sensors attempting to transmit at exactly the same time) occur during a transmission.
On the other hand, such a method entails the possibility of false positives; however, one can have a false positive probability as low as desired. 
As we will see in more detail when discussing the numerical results, such a disadvantage from the communication point of view does not interfere with the proper functionality of the system. 
In contrast, false negatives---resulting from a collision if ALOHA were adopted, for example---would substantially degrade the functionality of the system.

For $\mathcal{S}_3$, a possible approach in terms of periodic communication would be as follows. 
We could deploy three sensor nodes so that the specific position of any drone can be obtained by triangulation. 
Thus, each of the three sensor nodes would periodically send the distance between itself and the drones. 
In this way, the decision-maker would be able to estimate the position of each drone and, if necessary, send a return command. 
\subsection{Results}
We will now investigate the behavior of the proposed system. 
The impact of the determination of the semantic-functional states associated with different performance constraints $\delta_{\text{violation}}$ and $\delta_{\text{time}}$ can be assessed as follows.
Let us assume a total monitoring area of 60 m $\times$ 60 m, resulting in nine cells of 20 m  $\times$ 20 m.
In this scenario, we consider the trade-offs involving the use of implicit/explicit communications, measured by $\Theta[k]$, and the performance $\mathcal{V}_{\text{violation}}$ and $\mathcal{V}_{\text{time}}$ that evaluate coverage.

\begin{figure*}
    \center
    \subfigure[{Short-term walk for $\mathcal{S}_1$}]{
    \includegraphics[width=5cm]{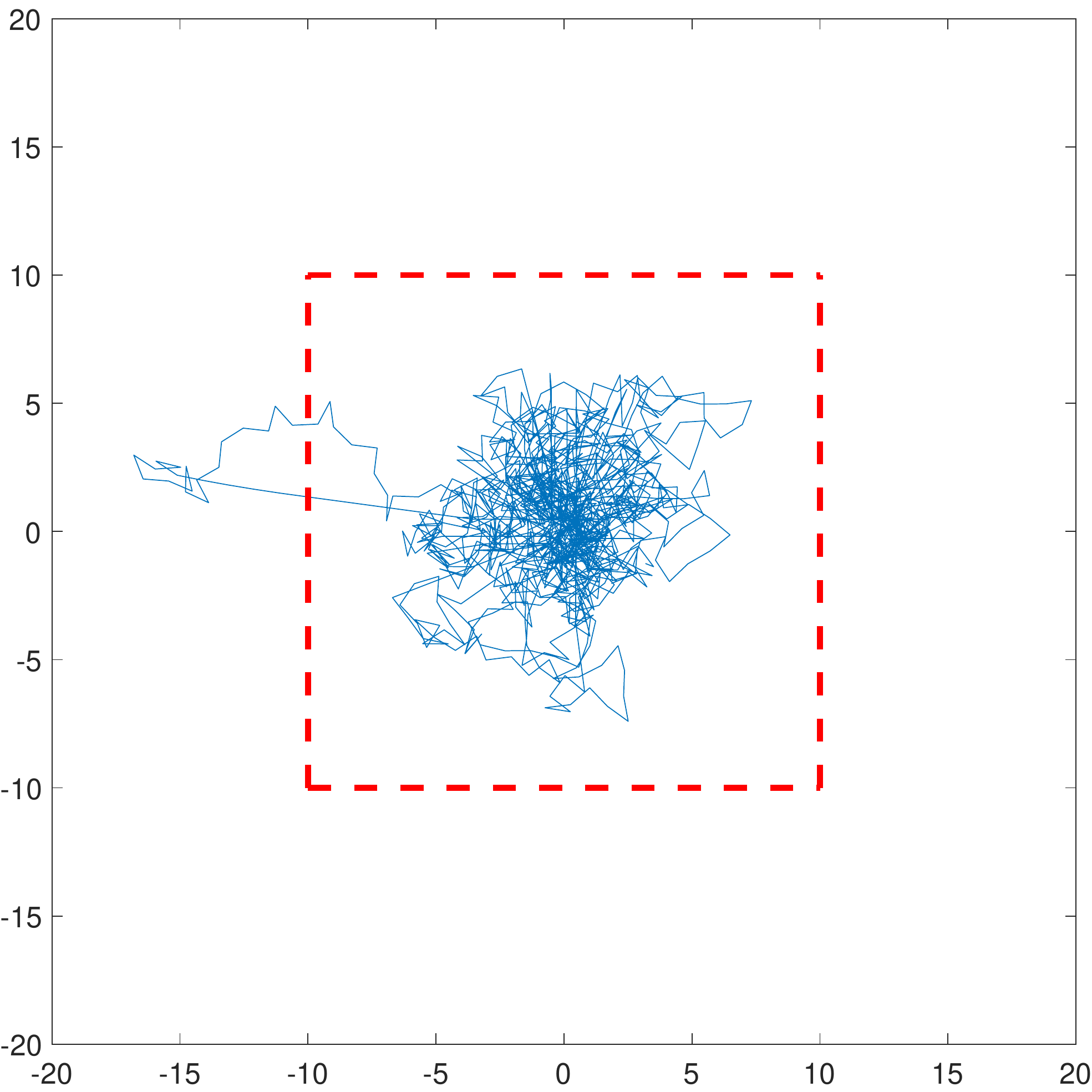}}
    \qquad
    \subfigure[{Short-term walk for $\mathcal{S}_2$}]{
    \includegraphics[width=5cm]{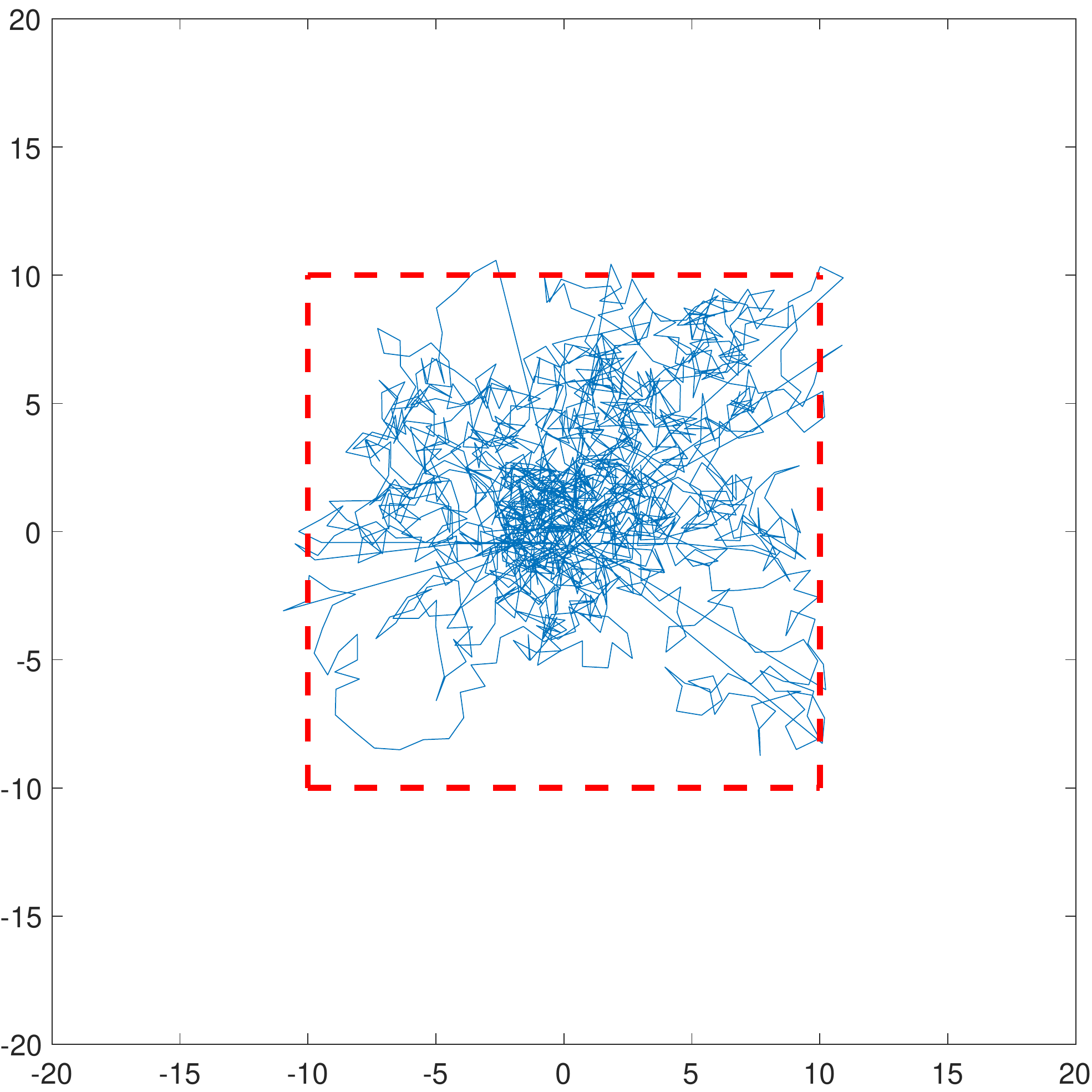}}
    \qquad
    \subfigure[{Short-term walk for $\mathcal{S}_3$}]{\includegraphics[width=5cm]{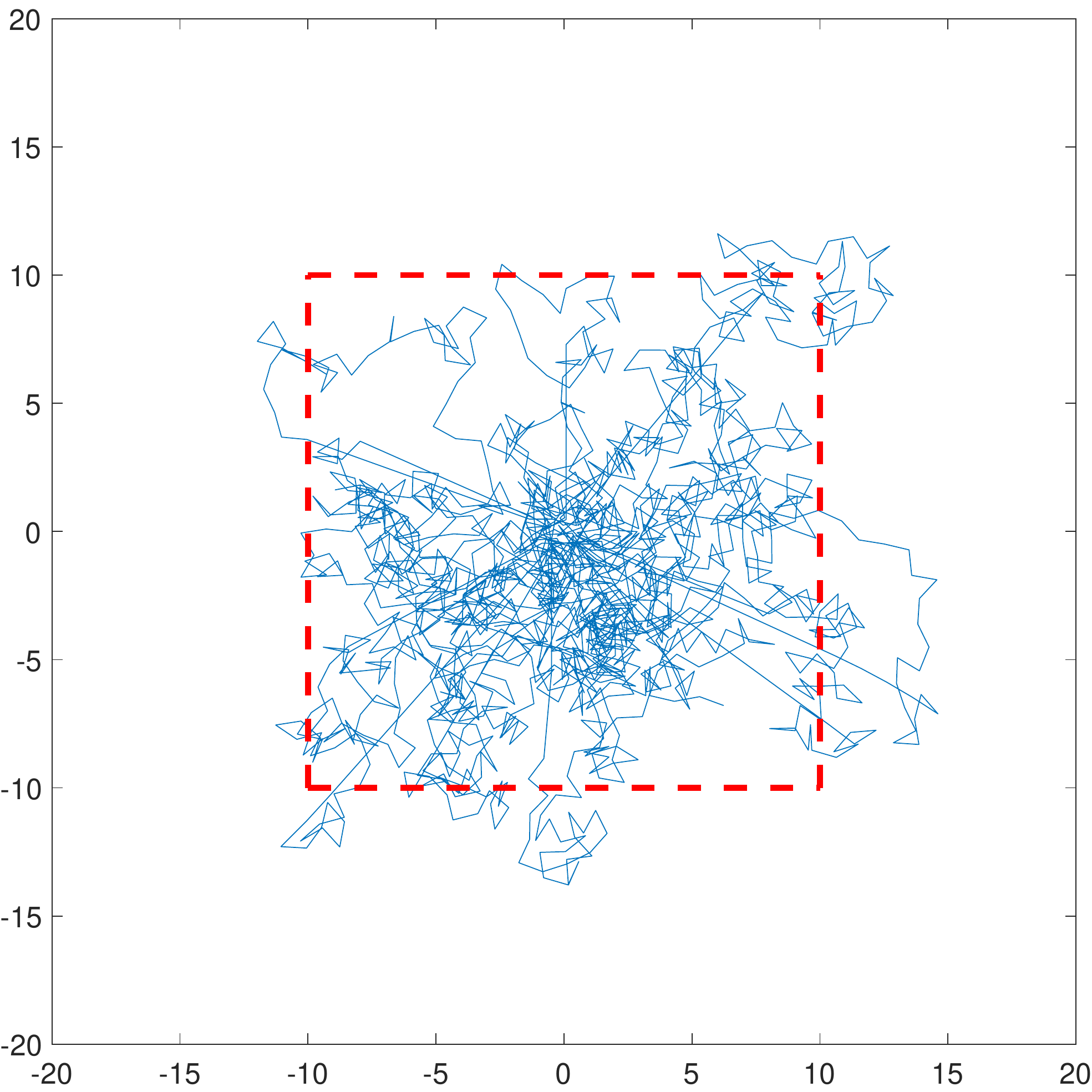}}
    \\
    \subfigure[{Long-term walk for $\mathcal{S}_1$}]{
    \includegraphics[width=5cm]{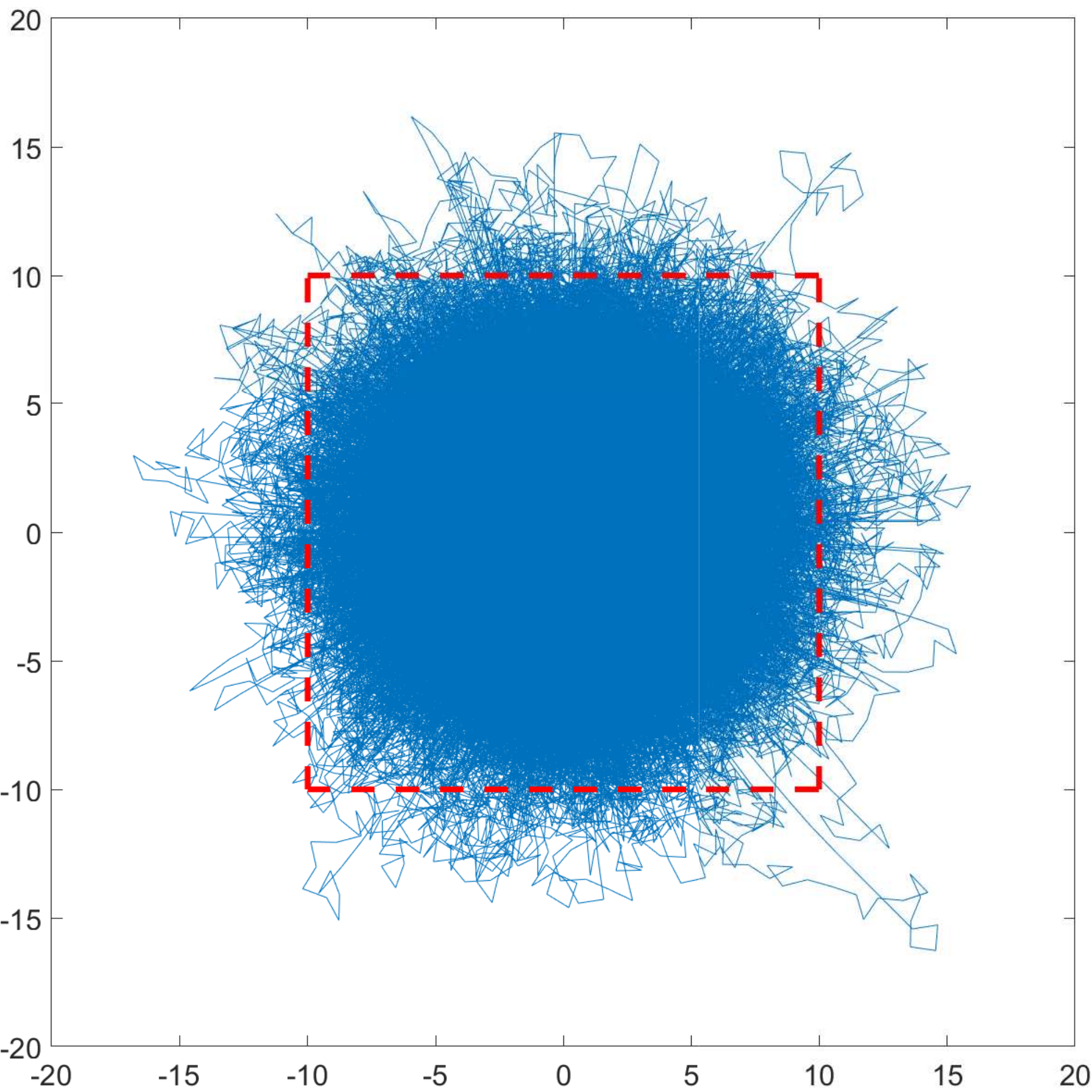}}
    \qquad
    \subfigure[{Long-term walk for $\mathcal{S}_2$}]{
    \includegraphics[width=5cm]{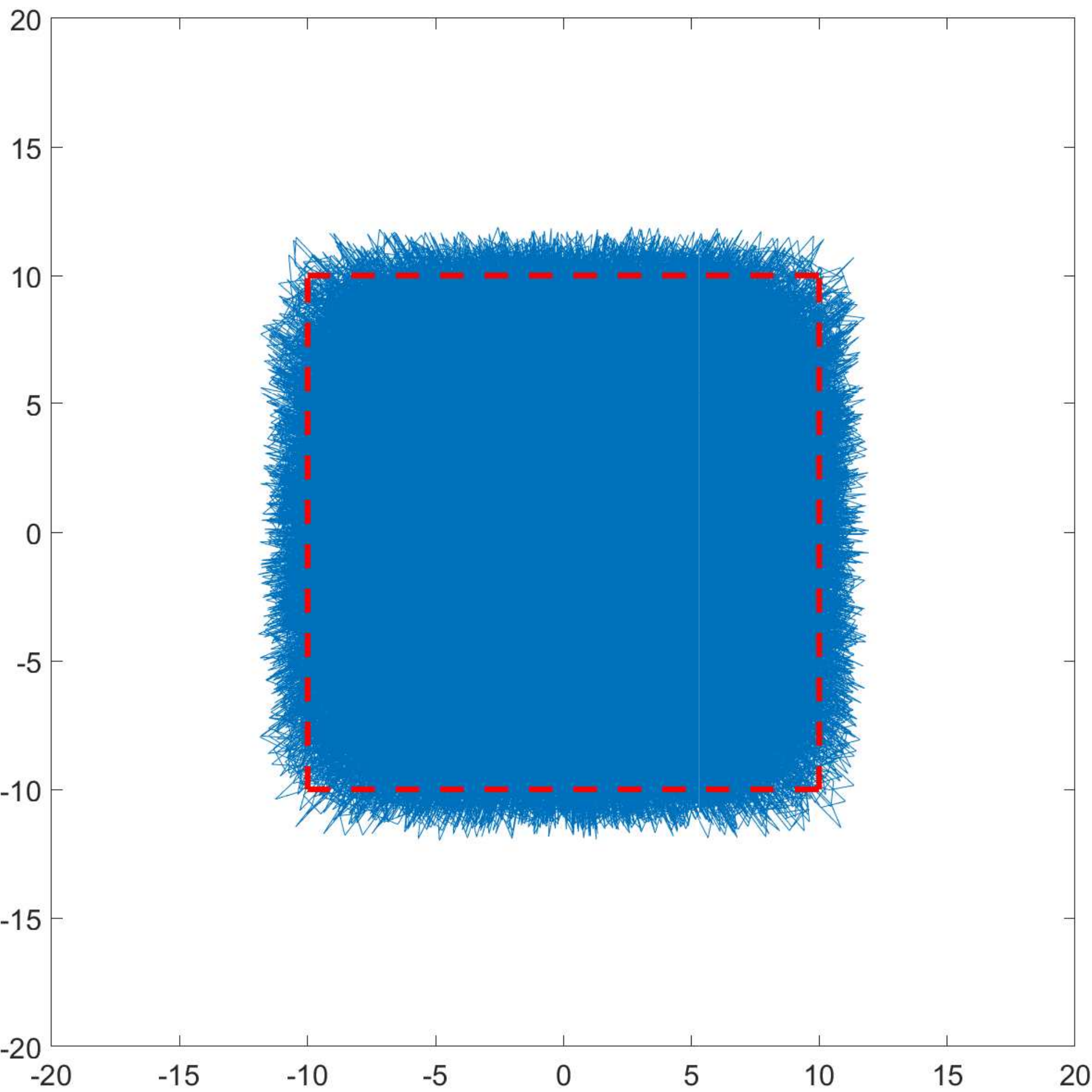}}
    \qquad
    \subfigure[{Long-term walk for $\mathcal{S}_3$}]{\includegraphics[width=5cm]{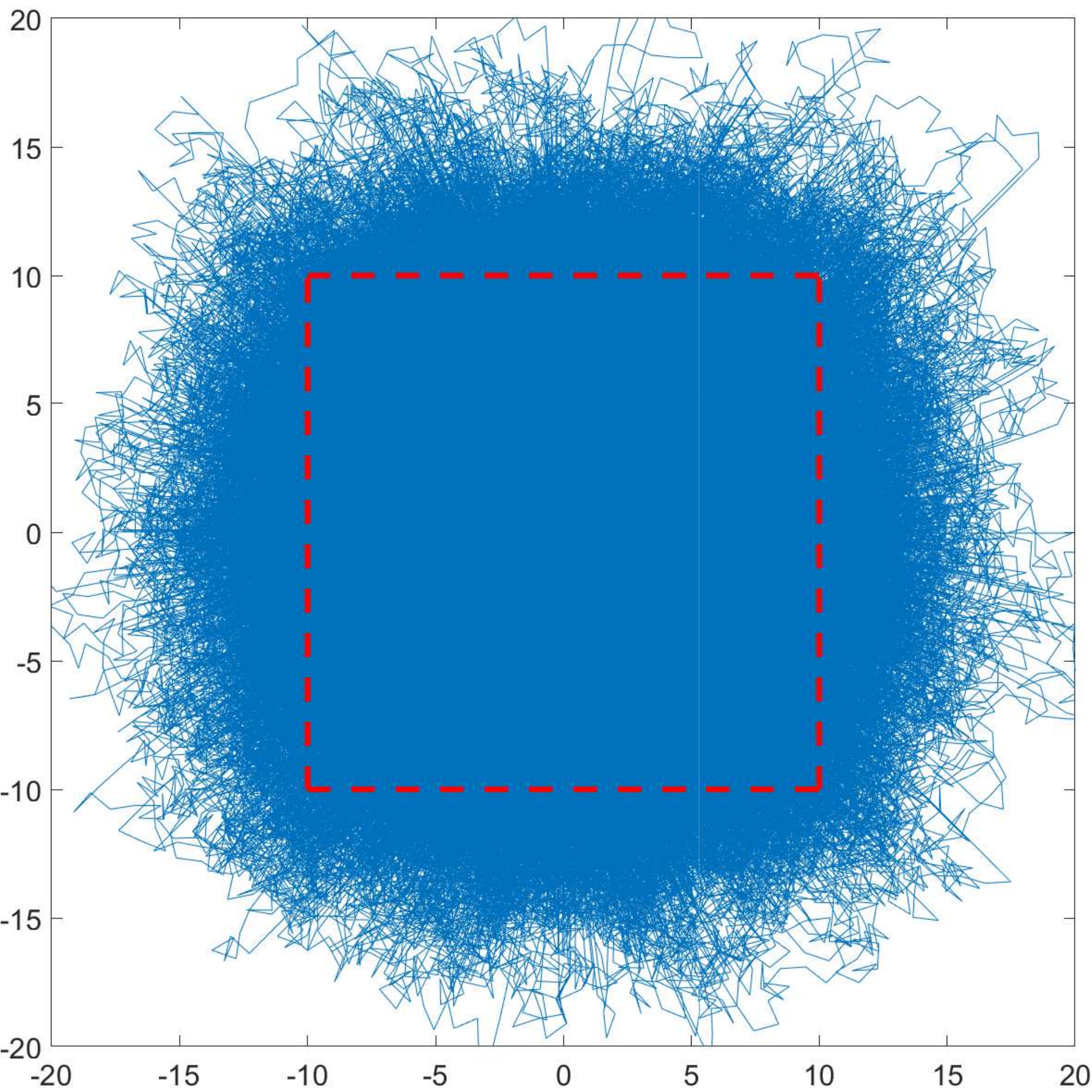}}
    \caption{Comparison between the semantic-functional approaches (a) and (d) $\mathcal{S}_1$ with $\mathbb{E}[\mathcal{V}_{\text{time}}]=71.44$ minutes and $\mathbb{E}[\mathcal{V}_{\text{violation}}]=0.35\%$, (b) and (e) $\mathcal{S}_2$ with $\mathbb{E}[\mathcal{V}_{\text{time}}]=9.04$ minutes and $\mathbb{E}[\mathcal{V}_{\text{violation}}]=1.08\%$, and (c) and (f) $\mathcal{S}_3$ with $\mathbb{E}[\mathcal{V}_{\text{time}}]=7.69$ minutes and $\mathbb{E}[\mathcal{V}_{\text{violation}}]=7.55\%$.}
    \label{fig:RW1}%
\end{figure*}

We will now turn our attention to the minimization of the use of explicit communication $\Theta[k]$. 
Given the three cases discussed in \cref{ssec:SFC cases}, we have $\Theta_1[k] = 0$, $0 < \mathbb{E}[\Theta_2[k]] < 1$, and $\Theta_3[k] = 1$ for $\mathcal{S}_1$, $\mathcal{S}_2$, and $\mathcal{S}_3$, respectively. 
Therefore, we can say that the composition of semantic-functional states ($\mathcal{S}_1$, $\mathcal{S}_2$, and $\mathcal{S}_3$) defines $\mathbb{E}[\Theta[k]]$.
Regarding $\mathcal{S}_1$, it is noteworthy that (i) if the frequency of the return command is high, the drones will tend to stay too close to the center in addition to overloading the communication system; and (ii) if the time between two subsequent return commands for the same drone are too long, it will tend to not remain inside its cell.
It is assumed that the discrete interval $k$ lasts $T_{\text{k}}$ seconds.
 
The communication for the second scenario is addressed as follows. 
By setting the time dependence as $d=1$ and transmitting according to the \ac{SFC} technique of \cite{silva2022semantic} only when $s_k \neq s_{k-1}$, we define transmission function $m$. 
In this way, the sensing channel remains idle whenever $s_k = s_{k-1}$.
Furthermore, the decision-maker can always estimate correctly the semantic-functional state at instant $k$ by setting $\hat{s}_k=\hat{s}_{k-1}$ whenever no sensor has transmitted.
The same value of transmission energy expenditure $\mathcal{E}$ is adopted for the three cases.
Furthermore, we assume that a transmission, starting from both the sensor nodes and the decision-maker, takes $T_{\text{k}}=0.1$ seconds for the three cases.
Regarding the communication between the sensors and the decision-maker, note that $\textup{Pr}[\hat{s}_k \neq s_k]$ is inversely proportional to $\mathcal{E}$ exclusively for $\mathcal{S}_2$, whereas $\mathcal{S}_3$ requires $\mathcal{E}$ to be greater than a minimum operating value.
This minimum value of $\mathcal{E}$ for $\mathcal{S}_3$ can be obtained through the Shannon's channel capacity.
We obtain $\mathbb{E}[\Theta_2[k]]=0.0577$ per second and $\textup{Pr}[\hat{s}_k \neq s_k]=2 \times 10^{-3}$ for $\mathcal{S}_2$ with a given set of drone random walk parameters.

The third hypothetical scenario has only three sensor nodes that constantly report their readings to the decision-maker as briefly discussed above. 
In this approach, only one sensor node can transmit at a time, and the decision-maker can know the exact position of a drone through triangulation. 
The sensor nodes update their readings to the decision-maker using the \ac{TDMA} technique. 
A \ac{TDMA} slot has $T_{\text{k}}$ seconds---maintaining consistency with proposed solutions---and can convey information (distance between drone and sensor node) about only one drone. 
Hence, the drone position is updated at the decision-maker side each $27 T_{\text{k}}$. 
We further assumed a random delay in the transmission of the return command; this delay reflects the time required to process the triangulation.
Note that the decision-maker notices a drone outside its respective area with a delay of at least $3 T_{\text{k}}$ and a maximum of $27 T_{\text{k}}$, excluding the triangulation processing time. 
Accordingly, we assume that the time to decision-maker detects a drone outside its respective area follows a uniform distribution between $3 T_{\text{k}}$ and $27 T_{\text{k}}$ seconds.

Figure \ref{fig:RW1} outlines the results for these three hypothetical scenarios. 
The figure presents the systems (a) and (d) without using explicit communication, i.e., $\mathcal{S}_1$; (b) and (e) minimum explicit communication with the \ac{SFC} representing $\mathcal{S}_2$; and (c) and (f) $\mathcal{S}_3$, which has maximum explicit communication.
Furthermore, (a), (b), and (c) show the path of a drone after a short period of $2 T_{\text{k}} \cdot 10^{3} $ seconds, while (d), (e), and (f) show the path of a drone after a long period of $2 T_{\text{k}} \cdot 10^{6}$ seconds.
Without explicit communication (i.e., the decision-maker operates blindly), we obtain $\mathbb{E}[\mathcal{V}_{\text{time}}]=71.44$ minutes and $\mathbb{E}[\mathcal{V}_{\text{violation}}]=0.35\%$ for an interval between two consecutive return commands uniformly distributed between 2 and 5 seconds. 
For $\mathcal{S}_2$, we have $\mathbb{E}[\mathcal{V}_{\text{time}}]=9.04$ minutes and $\mathbb{E}[\mathcal{V}_{\text{violation}}]=1.08\%$. 
Lastly, for $\mathcal{S}_3$, which constantly accesses the channel, we have $\mathbb{E}[\mathcal{V}_{\text{time}}]=7.69$ minutes and $\mathbb{E}[\mathcal{V}_{\text{violation}}]=7.55\%$. 
Note that $\mathcal{S}_1$ is the approach that keeps the drones closest to the center, $\mathcal{S}_2$ provides the most uniform coverage, and $\mathcal{S}_3$ slightly outperforms $\mathcal{S}_2$ in average coverage time $\mathbb{E}[\mathcal{V}_{\text{time}}]$.

\begin{figure}
 \centering
 \includegraphics[width=1\columnwidth]{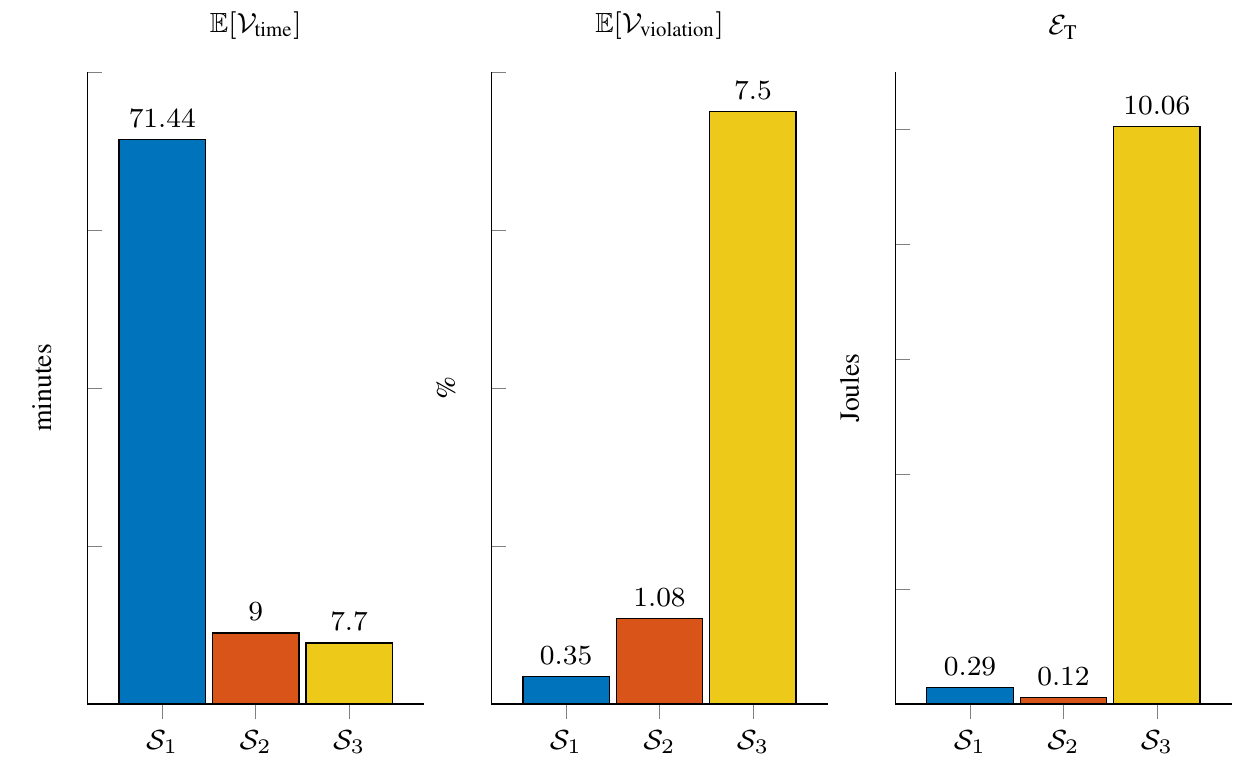}
 \caption{Comparison between the degrees of freedom to semantic-functional.}
  \label{fig:perf}
\end{figure}

Let $\mathcal{E}_{\text{T}}$ be the average energy spent on transmission for a subarea---we have included both the return command transmission and the sensor transmissions---per unit time normalized with respect to $\mathcal{E}$.
As mentioned, we assume that both a return command and a sensor transmission require $\mathcal{E}$ Joules. 
Thus, the calculation of $\mathcal{E}_{\text{T}}$ can be performed by adding the average number of sensor transmissions per second with the average number of return commands per second.
For instance, we have that average sensor transmission is zero and average return command per second is $0.2857$ (i.e, the inverse of the average time between the return commands sent to the same drone) for $\mathcal{S}_1$.

\Cref{fig:perf} compares the three degrees of freedom for the semantic-functional states adopted in this work.
Three evaluation parameters are presented: transmission energy expenditure $\mathcal{E}_{\text{T}}$, average coverage time $\mathbb{E}[\mathcal{V}_{\text{time}}]$, and percentage of violation (outside) of area $\mathbb{E}[\mathcal{V}_{\text{violation}}]$. 
Note that we are dealing with three distinct cyber domains (i.e., $\mathcal{S}_1$, $\mathcal{S}_2$, and $\mathcal{S}_3$) while the physical domain remains unchanged.
\Cref{fig:perf} shows the impact of the cyber domain on the physical one.
\section{Conclusions}
Semantic communications have attracted attention in recent years, possibly because of the rapid development of \ac{AI} methods that enable extracting semantic features from the data.
In this context, this work provides a new approach by leveraging implicit communication based on semantic knowledge in the design of \acp{CPS}.
Specifically, our main target was not to maximize the efficiency of the radio resources based on contextual knowledge, but rather to minimize the use of such resources while the \ac{CPS} under consideration can still operate within the desirable performance range.
With respect to the literature, we presented the formulation of what we call semantic-functional communication and successfully tested it in a \ac{CPS} constituted by a swarm of nine drones that jointly perform remote sensing over a certain region.
The promising results we obtained set the path for a more organic collaboration between control and communication engineers, both to determine the fundamental limits of the operation of cyber-physical systems, and to build more effective systems that use as little resources as possible to perform their particular tasks.

\bibliographystyle{IEEEtran}
\bibliography{IEEEabrv,References}


\end{document}

%% file: acronyms.tex
\DeclareAcronym{FDM}{
	short = FDM,
	long  = frequency-division multiplexing,
	tag = acrs,
}
\DeclareAcronym{OFDM}{
	short = OFDM,
	long  = orthogonal frequency-division multiplexing,
	tag = acrs,
}
\DeclareAcronym{CFR}{
	short = CFR,
	long  = channel frequency response,
	tag = acrs,
}
\DeclareAcronym{OFDMA}{
	short = OFDMA,
	long  = orthogonal frequency-division multiple access,
	tag = acrs,
}

\DeclareAcronym{O-OFDM}{
	short = O-OFDM,
	long  = optical orthogonal frequency division multiplexing,
	tag = acrs,
}

\DeclareAcronym{CD}{
	short = CD,
	long  = chromatic dispersion,
	tag = acrs,
}

\DeclareAcronym{PMD}{
	short = PMD,
	long  = polarization mode dispersion,
	tag = acrs,
}

\DeclareAcronym{3-D}{
	short = 3-D,
	long  = three-dimensional,
	tag = acrs,
}

\DeclareAcronym{2-D}{
	short = 2-D,
	long  = two-dimensional,
	tag = acrs,
}

\DeclareAcronym{4-D}{
	short = 4-D,
	long  = four-dimensional,
	tag = acrs,
}

\DeclareAcronym{ELU}{
	short = ELU,
	long  = exponential linear unit,
	tag = acrs,
}

\DeclareAcronym{GELU}{
	short = GELU,
	long  = Gaussian error linear unit,
	tag = acrs,
}

\DeclareAcronym{RELU}{
	short = RELU,
	long  = rectified linear unit,
	tag = acrs,
}

\DeclareAcronym{SELU}{
	short = SELU,
	long  = scales exponential linear unit,
	tag = acrs,
}
\DeclareAcronym{LTE}{
	short = LTE,
	long  = long term evolution,
	tag = acrs,
}
\DeclareAcronym{4G}{
	short = 4G,
	long  = fourth generation,
	tag = acrs,
}
\DeclareAcronym{5G}{
	short = 5G,
	long  = fifth generation,
	tag = acrs,
}
\DeclareAcronym{6G}{
	short = 6G,
	long  = sixth generation,
	tag = acrs,
}
\DeclareAcronym{SDMA}{
	short = SDMA,
	long  = space division multiple access,
	tag = acrs,
}
\DeclareAcronym{SDMA-OFDM}{
	short = SDMA-OFDM,
	long  = space division multiple access orthogonal frequency-division multiplexing,
	tag = acrs,
}
\DeclareAcronym{NOMA}{
	short = NOMA,
	long  = non-orthogonal multiple access,
	tag = acrs,
}
\DeclareAcronym{MIMO}{
	short = MIMO,
	long  = multiple-input-multiple-output,
	tag = acrs,
}
\DeclareAcronym{MIMO-NOMA}{
	short = MIMO-NOMA,
	long  = multiple-input-multiple-output non-orthogonal multiple access,
	tag = acrs,
}
\DeclareAcronym{CNN}{
	short = CNN,
	long  = convolutional neural network,
	tag = acrs,
}

\DeclareAcronym{NN}{
	short = NN,
	long  = neural network,
	tag = acrs,
}

\DeclareAcronym{LOS}{
	short = LOS,
	long  = line-of-sight,
	tag = acrs,
}
\DeclareAcronym{NLOS}{
	short = NLOS,
	long  = non-line-of-sight,
	tag = acrs,
}
\DeclareAcronym{WINNER}{
	short = WINNER,
	long  = Wireless World Initiative for New Radio,
	tag = acrs,
}
\DeclareAcronym{QuaDRiGa}{
	short = QuaDRiGa,
	long  = Quasi Deterministic Radio Channel Generator,
	tag = acrs,
}
\DeclareAcronym{MT}{
	short = MT,
	long  = mobile terminal,
	tag = acrs,
}
\DeclareAcronym{BS}{
	short = BS,
	long  = base station,
	tag = acrs,
}
\DeclareAcronym{NR}{
	short = NR,
	long  = new radio,
	tag = acrs,
}
\DeclareAcronym{AWGN}{
	short = AWGN,
	long  = additive white Gaussian noise,
	tag = acrs,
}
\DeclareAcronym{SNR}{
	short = SNR,
	long  = signal-to-noise ratio,
	tag = acrs,
}
\DeclareAcronym{ADAM}{
	short = ADAM,
	long  = Adaptive Moment Estimation,
	tag = acrs,
}
\DeclareAcronym{MSE}{
	short = MSE,
	long  = Mean Squared Error,
	tag = acrs,
 }
\DeclareAcronym{CPS}{
	short = CPS,
	long  = cyber-physical system,
	tag = acrs,
}
\DeclareAcronym{NCS}{
	short = NCS,
	long  = networked control system,
	tag = acrs,
}
\DeclareAcronym{AI}{
	short = AI,
	long  = Artificial Intelligence,
	tag = acrs,
}
\DeclareAcronym{SFC}{
	short = SFC,
	long  = semantic-functional communications,
	tag = acrs,
}
\DeclareAcronym{ETC}{
	short = ETC,
	long  = event-triggered control,
	tag = acrs,
}
\DeclareAcronym{TDMA}{
	short = TDMA,
	long  = Time-Division Multiple Access,
	tag = acrs,
}